\documentclass{ptephy_v1}
\usepackage{siunitx}
\preprintnumber{XXXX-XXXX} 

\usepackage{url}
\usepackage{textcomp}

\begin{document}

\title{Development of fast-response PPAC with strip-readout for heavy-ion beams}


\author[1,*]{Shutaro~Hanai}
\affil{Center for Nuclear Study, the University of Tokyo, Wako, Saitama 351-0198, Japan}

\author[2, 1]{Shinsuke~Ota}

\affil{Research Center for Nuclear Physics, Osaka University, Ibaraki, Osaka 567-0047, Japan}

\author[1]{Reiko~Kojima}

\author[1]{Shoichiro~Masuoka}

\author[3]{Masanori~Dozono}

\affil{Department of Physics, Kyoto University, Kyoto, Kyoto 606-8502, Japan}

\author[1]{Nobuaki~Imai}

\author[1]{Shin'ichiro~Michimasa}

\author[1, 4, 2]{Susumu~Shimoura}

\author[3]{Juzo~Zenihiro} 

\author[3]{Kento~Inaba}

\author[3, 4]{Yuto~Hijikata}

\affil{RIKEN Nishina Center for Accelerator-Based Science, Wako, Saitama 351-0198, Japan}

\author[5]{Ru~Longhi}

\author[1]{Ryo~Nakajima}

\affil{Instutute of Modern Physics, Chinese Academy of Sciences, Lanzhou, 730000, Gansu, China \email{hanai@cns.s.u-tokyo.ac.jp}}


\begin{abstract}%
A strip-readout parallel-plate avalanche counter (SR-PPAC) has been developed aiming at the high detection efficiency and good position resolution in high-intensity heavy-ion measurements. The performance was evaluated using 115~MeV/u $^{132}$Xe, 300~MeV/u $^{132}$Sn, and 300~MeV/u $^{48}$Ca beams. A detection efficiency beyond 99\% for these beams is achieved even at an incident beam intensity of 0.7 billion particles per second. The best position resolution achieved is 235~{\textmu}m (FWHM).
\end{abstract}

\subjectindex{xxxx, xxx}

\maketitle

\section{Introduction}

Radioactive isotope (RI) beams are powerful tools in experimental nuclear physics.
The intensity of RI beams provided by accelerator facilities such as RIBF in RIKEN~\cite{riken} and FRIB at Michigan State University~\cite{frib}, is improved year by year. The highest beam intensity is more than 10$^{6}$~counts per second and more exotic phenomena are now within our scope of study thanks to high-intensity RI beams.
To take full advantage of these advances, the detectors used in beam experiments must have the stability to maintain high performance under the high-counting conditions.

Position-sensitive detectors play important roles not only in measuring the beam profile but also in measuring the beam momentum at a momentum-dispersive focal plane on an event-by-event basis.
For in-flight position measurement of heavy ions, gas-filled detectors are widely used as beam-line detectors.
A parallel-plate avalanche counter (PPAC) operated with low-pressure gas contains only counter gas and thin metal-coated electrodes, and its material thickness is quite thin and uniform. 
Delay-line readout of the position-sensitive PPAC~\cite{dl1,dl2} is commonly used, and in particular, the standard for most experiments performed in RIBF.
The advantage of delay-line readout is that the number of readout channels is as few as five per one set of position determination (in vertical and horizonral).
In the delay-line readout PPAC (DL-PPAC), each strip is connected to the delay line and the position of a particle is proportional to the time difference between the signals read from the two ends of the delay line.
Since the maximum delay time is approximately 100~ns in the standard detectors at RIBF~\cite{dl2}, the non-negligible signal pileup occurs and results in lower detection efficiency for high-intensity beams such as 10$^{6}$~cps (counts per second), especially widely spread beams like at the momentum-dispersive focal plane.

PPAC itself has an inherently quick response time because they amplify the electrons drifting between two electrodes a few millimeters apart and the drift time of electrons is within 30~ns for the typical drift lentgh of 4~mm in the standard DL-PPAC at RIBF.
Therefore, by reading signals from each strip electrode individually without using delay lines, the reduction in detection efficiency under high counting can be mitigated.
In this paper, the development of the strip-readout type PPAC (SR-PPAC) is described. 
The goal of the development is to achieve greater than 99\% detection efficiency even for the high-intensity heavy-ion beam near 10$^{6}$~cps, while keeping the typical position and timing resolution obtained by DL-PPACs, namely a few hundreds~{\textmu}m and a few hundreds ps, respectively.

The design of SR-PPAC is presented in Sec.~2.
In Sec.~3 and 4, the performances evaluated in two experiments are described.
The summary and conclusion are given in Sec.~5.

\section{Design for strip-readout method}

To achieve the required resolutions and high-counting rate simultaneously, the charge of each strip  should be collected with a good charge resolution for a position determination within a strip as described below. Furthermore the charge collection and conversion should be fast enough. 
In general, the Charge-to-Digital Converter is adopted for obtaining charge information, but the conversion time limits the counting rate of the whole system. Pulse shape analysis of fast sampling signals is an option for an ultra-fast conversion technique. However, the cost of such conventional fast digitizers is still high.
For these reasons, the Time-over-Threshold (ToT) method~\cite{tot} has been used in this development. The ToT method is a technique that extracts the pulse width at a certain threshold voltage, which is a monotonically increasing function of a pulse height or charge. The maximum output time duration is adjusted to be several tens of nanoseconds optimizing the pulse height threshold in an experiment.

In this section, the components and the configuration of the SR-PPAC including the readout electronics and the analysis method for the position extraction are described.

\subsection{Configuration of SR-PPAC}

The SR-PPAC is developed by remodeling the DL-PPAC. Table~1 shows  the configurations of SR-PPACs. Two types of SR-PPAC were constructed. One is called a prototype for element development, and the other is called a standard type with a larger sensitive area which can cover the typical beam distribution at a dispersive focal plane. The electrodes and the layout is the same as a typical delay-line PPAC~\cite{dl2} used in BigRIPS, RIKEN. 

\begin{table}[!h]
\caption{Configuration of SR-PPAC}
\label{table_example}
\centering
\begin{tabular}{c||c|c}
\hline
& Prototype & Standard \\ 
\hline
Sensitive area [mm$^{2}$] &	150 (X) $\times$ 150 (Y) & 240 (X) $\times$ 150 (Y)  \\
\hline
Gap between anode and cathode [mm] &	4 & 4.3  \\
\hline
Strip width [mm] & 2.57 (X, Y)  & 2.55 (X), 2.58 (Y) \\
\hline
Number of strips (channels)& 58 (X, Y)  & 94 (X), 58 (Y) \\
\hline
\end{tabular}
\end{table}

\begin{figure}[!h]
\begin{center}
\includegraphics[width=\hsize,clip]{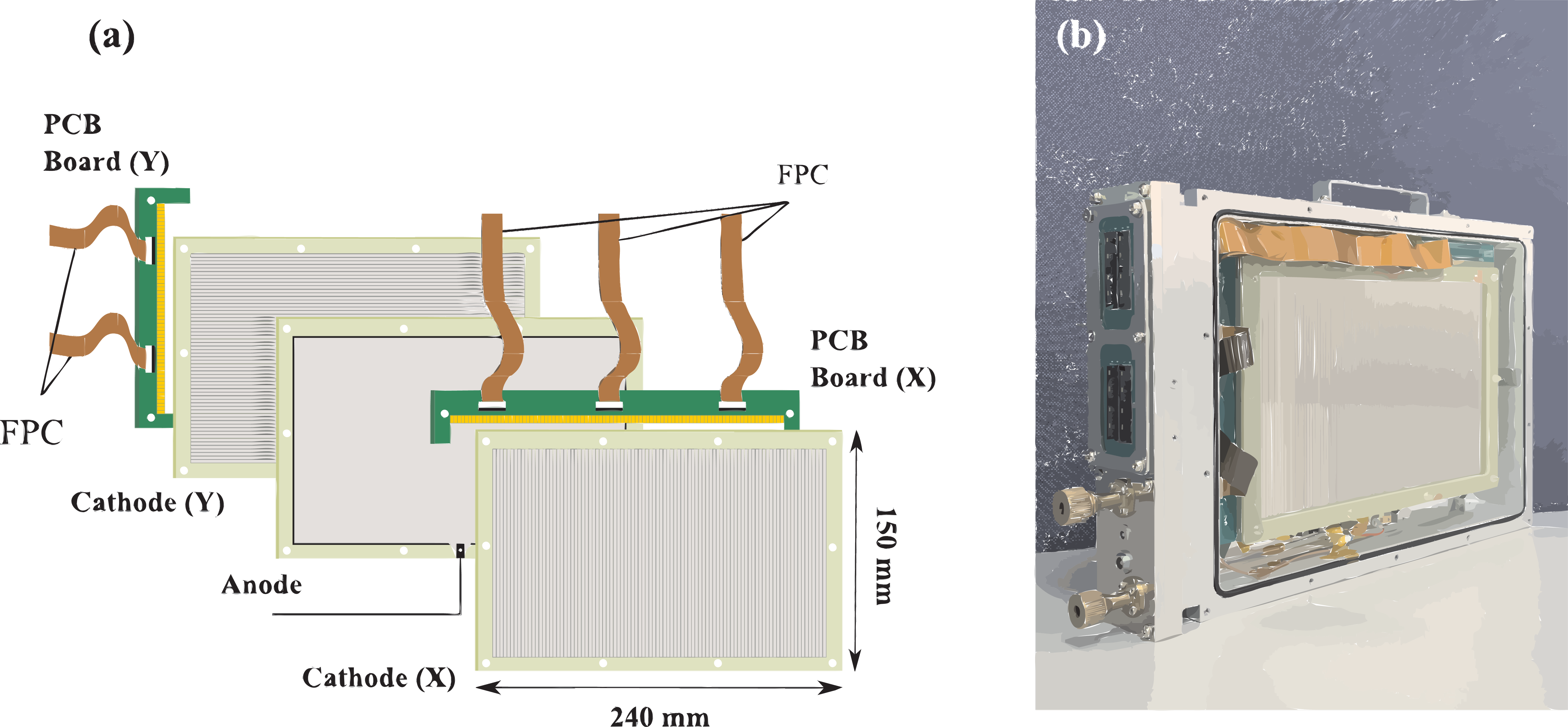}
\caption{\label{fig.structure} (a) Exploded view of the inner structure of a Standard-type SR-PPAC. (b) Photograph of a Standard-type SR-PPAC without the entrance window.}
\end{center}
\end{figure}

 Figure~\ref{fig.structure} (a) shows a schematic view of the standard SR-PPAC's inner structure. Newly-designed printed circuit boards (PCB) for readout from each strip are attached to the cathode foil such that metal-striped leads on the surface of the PCB are in contact with the cathode strip. Connectors for flexible printed circuits (FPC) instead of delay-line readout are mounted on the readout PCB for dealing with up to 94 ch and 58 ch for horizontal (X) and vertical (Y) directions perpendicular to the beam direction, respectively.  
 
 A signal feed-through PCB was also designed to extract the signals from the gas volume of the SR-PPAC to the outside.  
The feed-through PCB has surface-mounted connectors on both sides and the connectors are electrically connected using the pad-on-via technique ensuring this feed-through PCB holds vacuum. The standard-type SR-PPAC has eight surface-mount connectors (FH39A-67S-0.3HW) inside and four surface-mount connectors (QSE040-01-F-D) outside.

 The photograph in Fig.~\ref{fig.structure} (b) shows an external view of standard-type SR-PPAC. An entrance window is removed so that the inside is visible. On the left side of the detector, there are a feed-through PCB, gas inlet and outlet, and connectors for bias voltage and the signal from the anode electrode.
 
\begin{figure}[!h]
\begin{center}
\includegraphics[width=\hsize,clip]{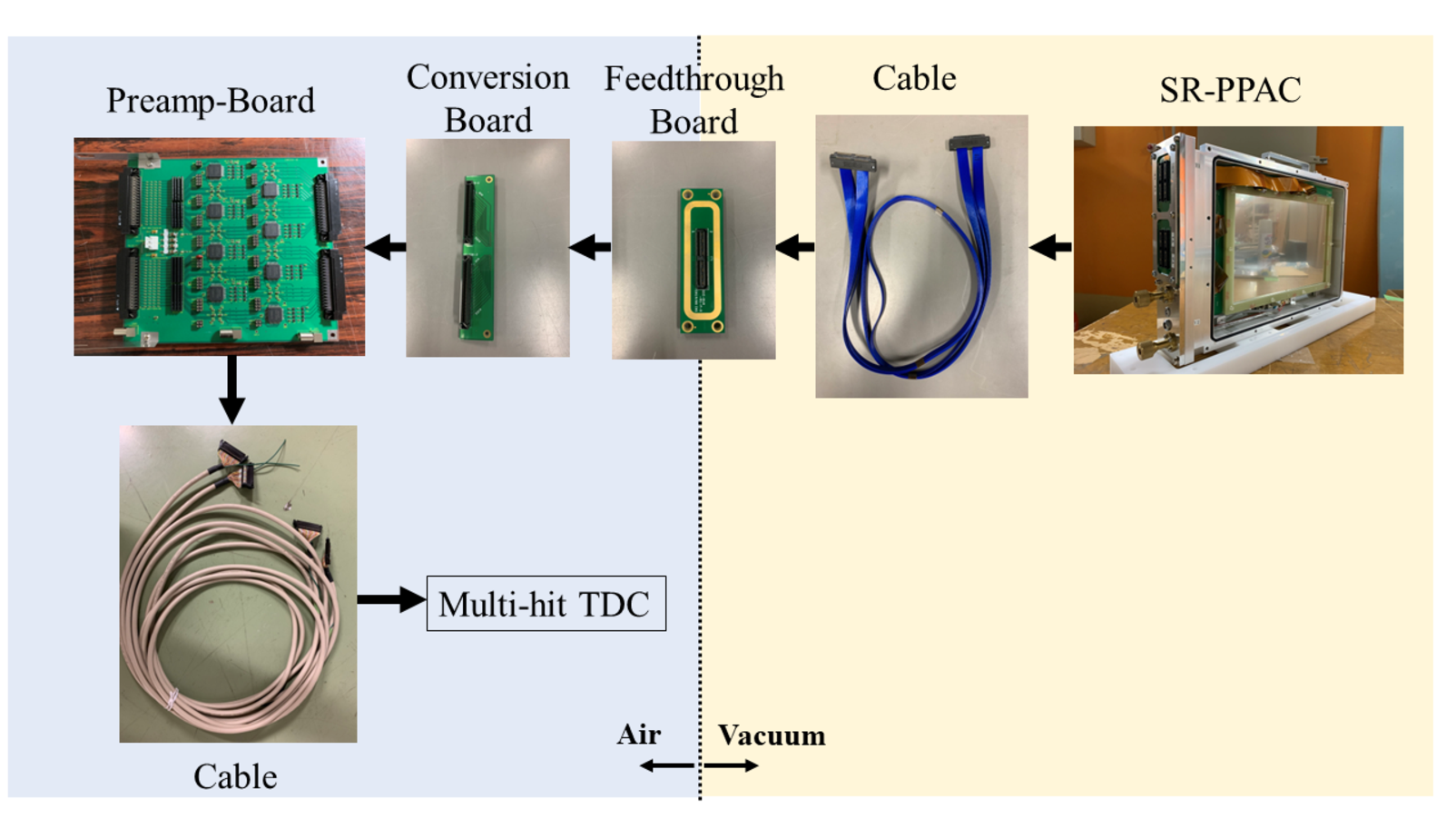}
\caption{\label{fig.trans}Typical signal transmission of SR-PPAC. The arrows show the flow of signal from the SR-PPAC to a TDC device. The feedthrough board and the conversion board were newly designed in this development. The components in the right side of the black botted line can be put in vacuum.}
\end{center}
\end{figure}

One or two SR-PPACs are typically located in large vacuum chambers. Typical signal transmission from SR-PPAC to the outside of the vacuum chamber is shown in Fig.~\ref{fig.trans}. Electrical signals are transmitted to preamplifier-shaper-discriminator (ASD) boards REPIC RPA-132, products of HAYASHI-REPIC Co., Ltd~\cite{repic} through a feed-through flange which is mounted on the vacuum chamber at each detector site. The specification of the ASD board is provided in Table~2.
 A pitch convertion board was designed to adapt the ASD board to the cable connector.
 The analog signals are converted to logic signals by the discriminator using the ToT method. The leading and trailing edges are acquired by multihit-TDC (CAEN V1190). The threshold voltage for ToT is adjusted considering the noise level  in the experimental site.

\begin{table}[t]
\caption{Specification of ASD board~\cite{repic}}
\label{table_example}
\centering
\begin{tabular}{c|c}
\hline
Number of channels & 64 ch \\ 
\hline
Input charge & $\pm$ 0.1 pC \\
\hline
Analog output & GAIN 0.8 V/pC \\
& Time constant of integrator 16 ns  \\
\hline
Threshold control voltage & - 0.6 V $\sim$ + 0.6 V \\
\hline
Size & 160 mm x 180 mm \\
\hline
\end{tabular}
\end{table}

\subsection{Method of position extraction}

A method was developed to derive the hit position using the amount of charge, in contrast to a DL-PPAC which uses timing information.

\begin{figure}[tbh]
\centering\includegraphics[width=\hsize,clip]{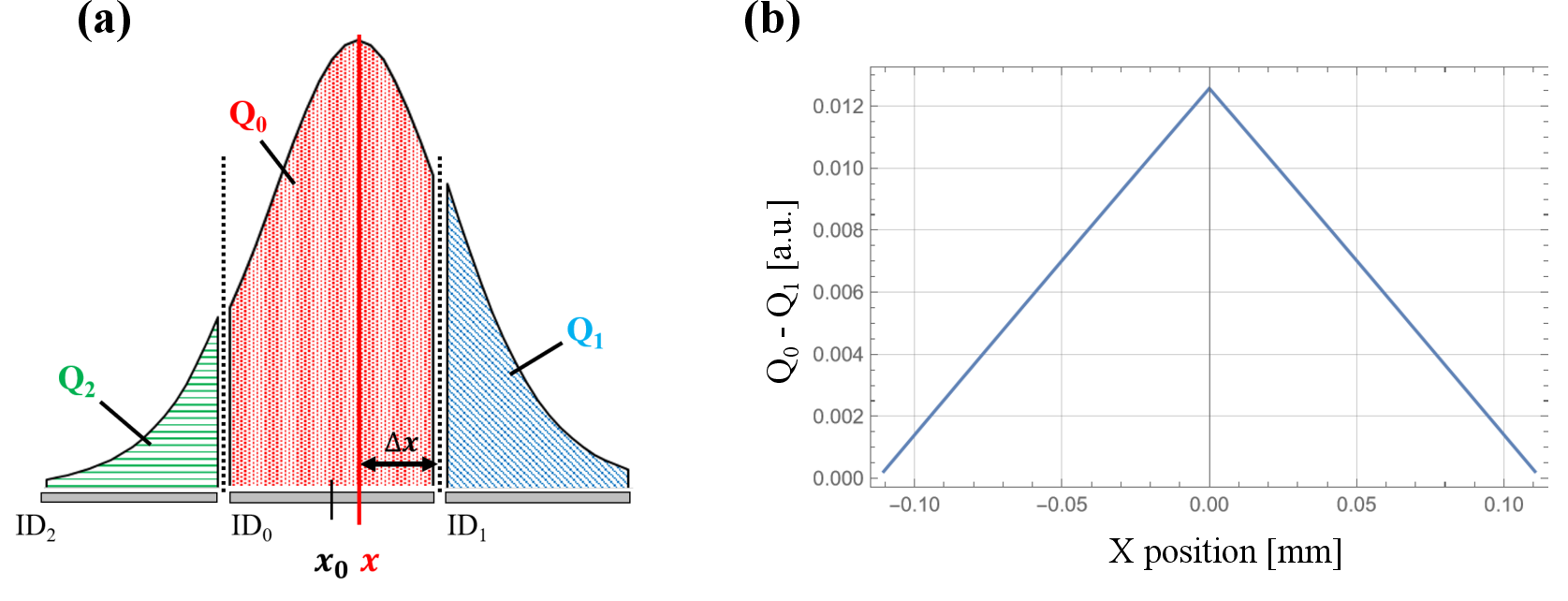}
\caption{\label{fig.estimate}(a) Illustration of charge distribution on the striped cathode induced by a point charge, ID$_{0}$, ID$_{1}$ and ID$_{2}$ represents the highest-, the second-highest, and the third-highest charged strip respectively. (b) Calculated correlation between charge difference and position. The $y$-axis represents charge difference between charge of ID$_{0}$ ($Q_{0}$) and ID$_{1}$ ($Q_{1}$).}
\end{figure}

When a particle enters a PPAC, electron and ion pairs are generated along the trajectory of the particle and those electrons immediately generate an avalanche drifting from the cathode to the anode.
Figure~\ref{fig.estimate} (a) shows an illustration of a distribution of the induced charge on the striped cathode. ID$_{0}$, ID$_{1}$ and ID$_{2}$ denote the index of the strip having the highest-, the second-highest, and the third-highest charged respectively. $Q_{0}$, $Q_{1}$, and $Q_{2}$ denote the amount of charge of ID$_{0}$, ID$_{1}$, and ID$_{2}$ respectively.
The difference of the charges induced on two adjacent strips has a correlation with the center of mass of the charge distribution. A beam position can be determined more precisely than the strip size of about 2.6~mm by using the charge difference between two strips ($\Delta q$). On the basis of the Townsend theory of electron avalanche\cite{townsent}, we estimated the charge distribution induced by a single point charge drifting from the cathode to the anode to illustrate this method. Figure~\ref{fig.estimate} (b) shows the charge difference of $Q_{0}$ and $Q_{1}$ as a function of  the position of the center of mass of the calculated distribution. Generally, this correlation between the position and charge difference is non-linear. 

\begin{figure}[tbh]
\centering\includegraphics[scale=0.5,clip]{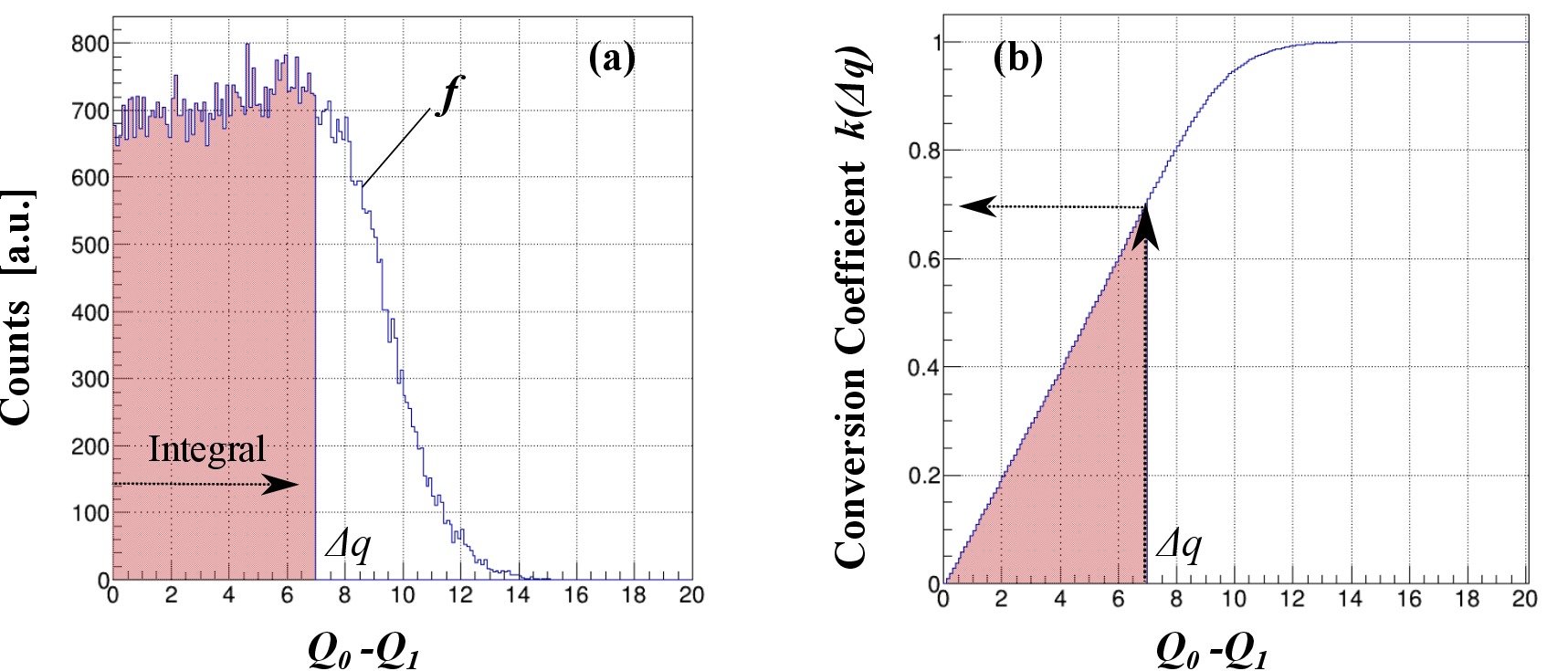}
\caption{\label{fig.dqdx}(a) A histogram of charge difference $Q_{0} - Q_{1}$ obtained from a measurement of all strips. The distribution is described as $f$ in Eq.~(2).  The integration of $f$ provides a conversion function as Eq.~(2). (b) A conversion coefficient denoted by $k$ as a function of  the charge difference. One $Q_{0} - Q_{1}$ value, $\Delta q$, corresponds to one $k$ value. A position will be given by multiplying $k$ with strip width $s$ as shown in Eq.~(1).}
\end{figure}

Assuming the uniform distribution of the particles hitting a strip, the integration of the distribution of $\Delta q$ can be written down as a single-valued function of the position distribution between the center of a strip and an edge of a strip. The distance between the beam position and the edge of a strip ($\Delta x$) can be calculated as follows:

\begin{equation}
\Delta x  =  \frac{s}{2} \times k (\Delta q)
\end{equation}

\begin{equation}
k  (\Delta q) = \frac{\int_{0}^{\Delta q} f(q) dq}{\int_{0}^{Q} f(q) dq}
\end{equation}

\noindent where $s$ is strip width, $k(\Delta q)$ is a conversion coefficient which is a function of  $\Delta q$, $Q$ is the maximum value of the charge difference, and $f$ is a distribution of charge differences between two strips, which is practically obtained from a measurement in the analysis.   Figure~\ref{fig.dqdx} (a) shows $\Delta q$ distribution obtained from the measured charge of all strips hit by particles. The conversion coefficient shown in Fig.~\ref{fig.dqdx} (b) provides a lookup table for position information from the charge information.
In the case that a beam arrives at the center of a strip, $\Delta q$  will be $Q$ or $\Delta x$ will be zero, and when a beam arrives at the edge of a strip, $\Delta q$ will be zero or $\Delta x$ will be the half size of a strip. Using Eq. (1) and (2), the beam position can be numerically obtained  without considering an exact form of the function to convert from charge to position.
Finally, the beam position will be given by the following equation:

\begin{equation}
x = x_{0} \pm ( \frac{s}{2} - \Delta x )  
\end{equation}

\noindent where $x_{0}$ is the central position of a strip which has the highest charge $Q_{0}$. The plus or minus sign in Eq. (3) is determined depending on which side of ID$_{0}$ a beam passed through, namely the position of the ID$_{1}$.

The charge difference used to deduce the position $x$ does not have to be $Q_{0} - Q_{1}$. For example, it is possible to use the charge difference between $Q_{0}$ and the charge on the left or right side of ID$_{0}$. The impact of the charge difference on position resolution is discussed in Sec.~3.4.

\section{Proof of the principle in the position extraction method for SR-PPAC}

\begin{figure}[!h]
\centering\includegraphics[scale=0.5]{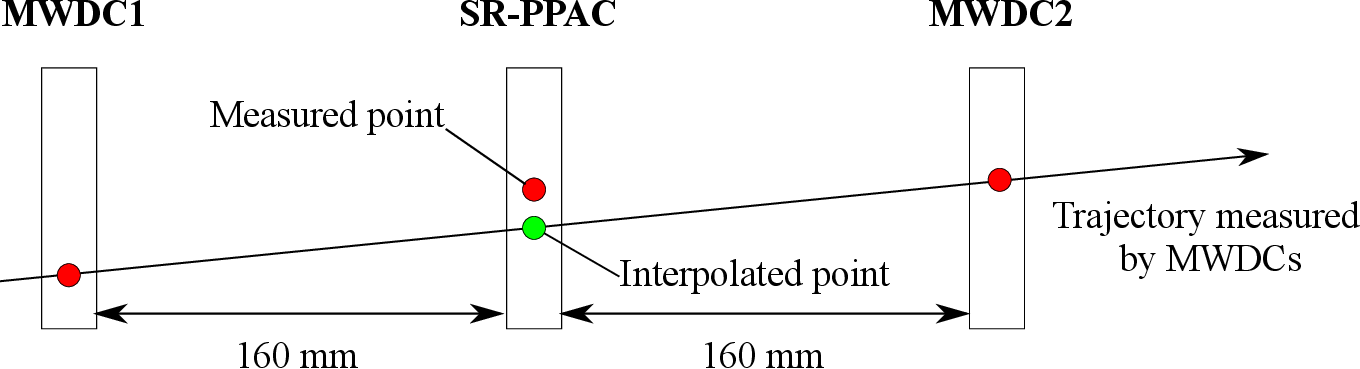}
\caption{\label{fig.setupHI}Illustration showing a setup of the experiment at HIMAC and how to deduce the position resolution using tracking by two MWDCs.} 
\end{figure}

 The first attempt to extract the charge from each strip and to evaluate the performance of the position extraction method was performed  by bombarding the prototype SR-PPAC with a beam of $^{132}$Xe at 115~MeV/u at HIMAC in Chiba, Japan (the experiment number 15H307). Counter gas was $i$-C$_{4}$H$_{10}$ and the two different pressure, 10~Torr and 5~Torr, were tested to evaluate the efficiency.
The experimental setup is shown in Fig.~\ref{fig.setupHI}. Two low-pressure multi-wire drift chambers (LP-MWDCs)~\cite{lpmwdc} were placed as reference detectors and the prototype SR-PPAC was placed between them.
We evaluated the detector performance, in particular detection efficiency, position resolution, and tracking efficiency.

Note that  notations of charge such as $Q_{0}$ and $Q_{1}$ described in Sec.~3 and Sec.~4 represent a time duration of a signal measured using ToT method and the units are nanoseconds.

\subsection{Detection efficiency}

\begin{figure}[!h]
\centering\includegraphics[scale=0.52]{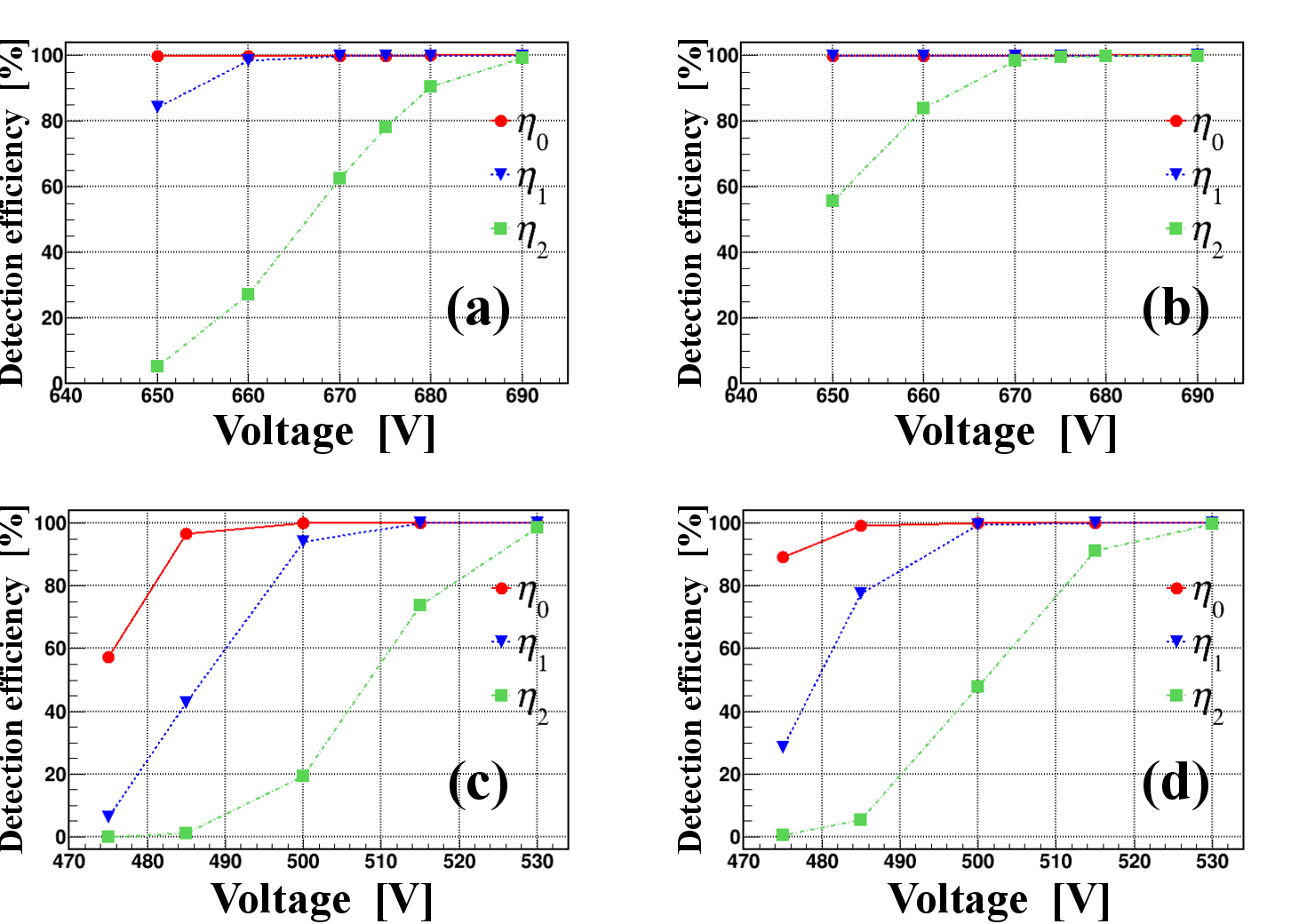}
\caption{\label{fig.effHI}Anode bias dependence of detection efficiency of SR-PPAC operated with the gas pressure of 10~Torr (a),(b) for Cathode X and Y respectively. (c) and (d) show 5~Torr for Cathode X, Y respectively. $\eta_{0}$, $\eta_{1}$, $\eta_{2}$ represents the detection efficiency of highest, the second-highest, and the third-highest charge respectively. }
\end{figure}

 The detection efficiency of SR-PPAC was defined by the following equations:

\begin{equation}
\eta_{xi} = \frac{N_{xi} \otimes N_{\mathrm{mwdc}}}{N_{\mathrm{mwdc}}}  \hspace{20pt} (i = 0, 1, 2)  
\end{equation}
\begin{equation}
\eta_{yi} = \frac{N_{yi} \otimes N_{\mathrm{mwdc}}}{N_{\mathrm{mwdc}}} \hspace{20pt} (i = 0, 1, 2)
\end{equation}

\noindent where $N_{xi}$, $N_{yi}$, and $N_{\mathrm{mwdc}}$ represent number of events in $x-$plane and $y-$plane in SR-PPAC, and  number of events in two LP-MWDCs, respectively. The index $i$ correnponds to ID$_{0}$, ID$_{1},$ and ID$_{2}$. For instance, $\eta_{x0}$ is the ratio of the number of hits in strips with the highest charge  to the number of hits in the MWDCs.  
Figure~\ref{fig.effHI} shows the detection efficiency as a function of applied anode bias. As mentioned in Sec.~2.1, $Q_{0}$ and $Q_{1}$ are necessary for position extraction in SR-PPAC. The detection efficiency of $Q_{1}$ is more than 99 \% at an anode bias of 670~V in 10~Torr and 515~V in 5~Torr. 

\begin{figure}[!h]
\centering\includegraphics[scale=0.45]{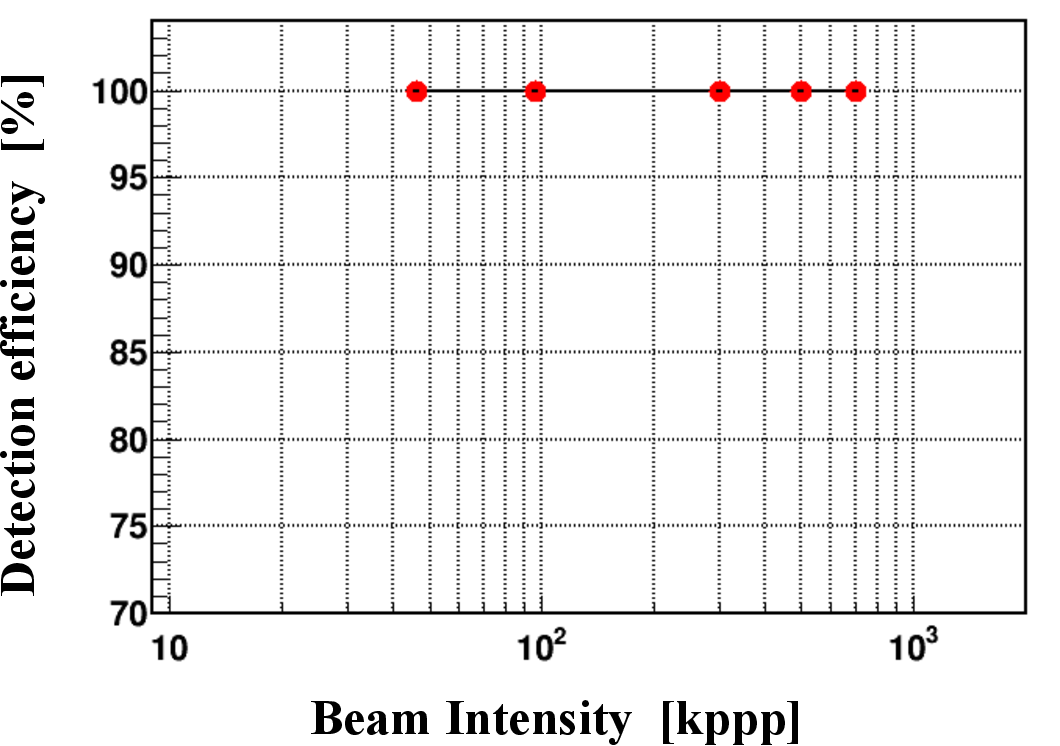}
\caption{\label{fig.effHII}Intensity dependence of detection efficiency ( $\eta_{x1}$ ). The applied bias to the anode was 525~V, and the gas pressure was 5~Torr.}
\end{figure}

 In Fig.~\ref{fig.effHII}, the detection efficiency of $Q_{1}$ (which requires the existence of the $Q_{0}$)  is plotted as a function of the beam intensity. The SR-PPAC was operated at 525~V in 5~Torr. Even for a beam of 700~kppp (ppp: particles per pulse), the detection efficiency is more than 99\%. Note that the beam size at the SR-PPAC was $\phi 20$~mm.
    
\subsection{Position resolution}

\begin{figure}[!h]
\centering\includegraphics[scale=0.7]{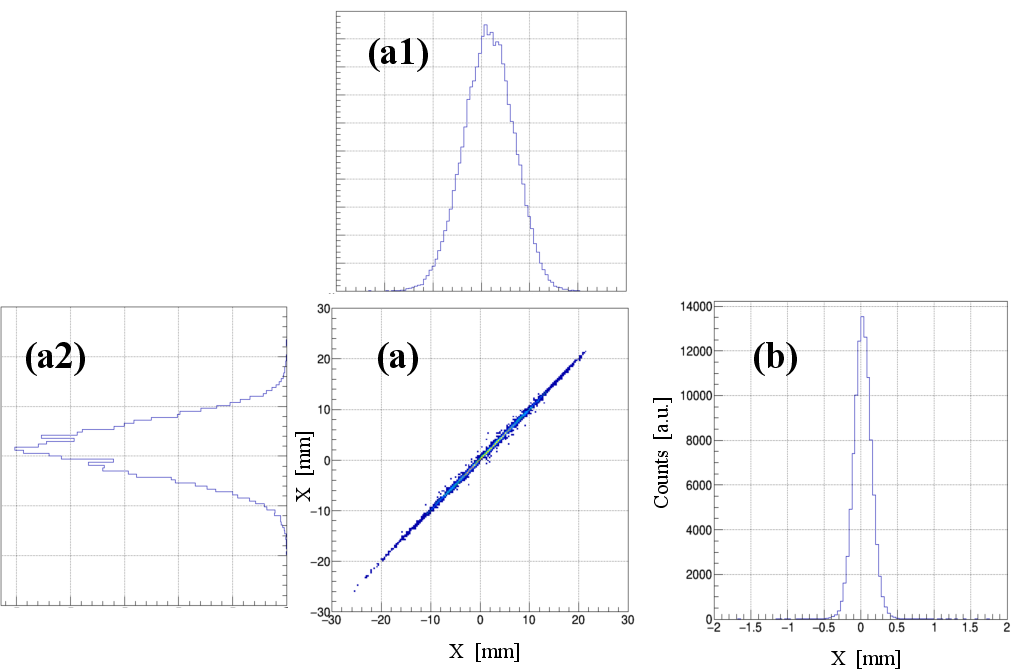}
\caption{\label{fig.posidist}(a) Two dimentional plot of interpolated position distribution using two MWDCs (a1) and position distribution detected by SR-PPAC (a2). (b) Difference between position in MWDCs and in SR-PPAC.}
\end{figure}

Position resolution of SR-PPAC was evaluated by using the difference of a beam position measured by SR-PPAC and a position interpolated by tracking using two LP-MWDCs as illustrated in Fig.~\ref{fig.setupHI}. Figure~\ref{fig.posidist} shows the distributions of the position measured by SR-PPAC, interpolated position, and their difference. The position resolution of SR-PPAC is calculated by the following equation:

\begin{equation}
\sigma_{\mathrm{srppac}} = \sqrt{ \sigma_{\mathrm{res}}^{2} - \sigma_{\mathrm{mwdc}}^{2} }
\end{equation}

\noindent where $\sigma_{\mathrm{srppac}}$, $\sigma_{\mathrm{mwdc}}$, and $\sigma_{\mathrm{res}}$ means the position resolution of SR-PPAC and LP-MWDC, and standard deviation of the distribution of the difference, respectively. $\sigma_{\mathrm{mwdc}}$ was deduced by using the average value of the unbiased variance of the position at each plane of LP-MWDC.

\begin{figure}[!h]
\centering\includegraphics[width=\hsize]{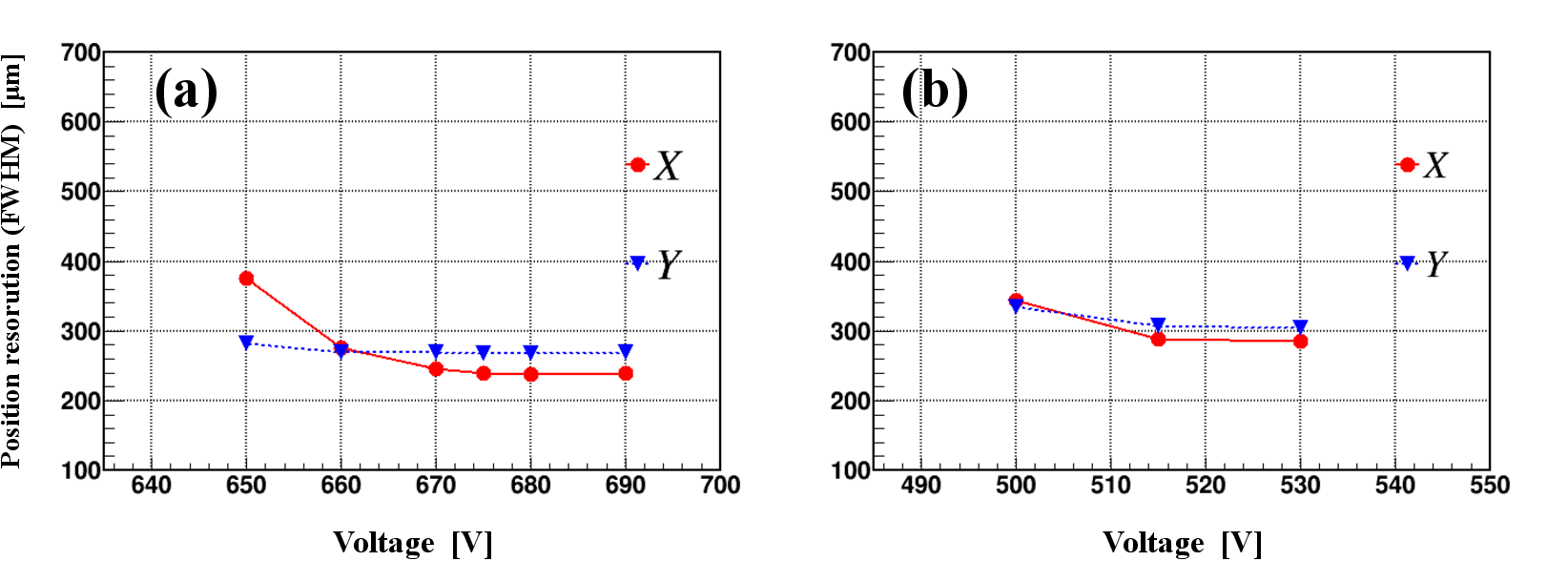}
\caption{\label{fig.posiHI} Position resolution of SR-PPAC. The gas pressure was 10~Torr (a) and 5~Torr (b).}
\end{figure}

\begin{figure}[tb]
\centering
\begin{minipage}{0.48\hsize}
\centering
\includegraphics[height=5cm,clip]{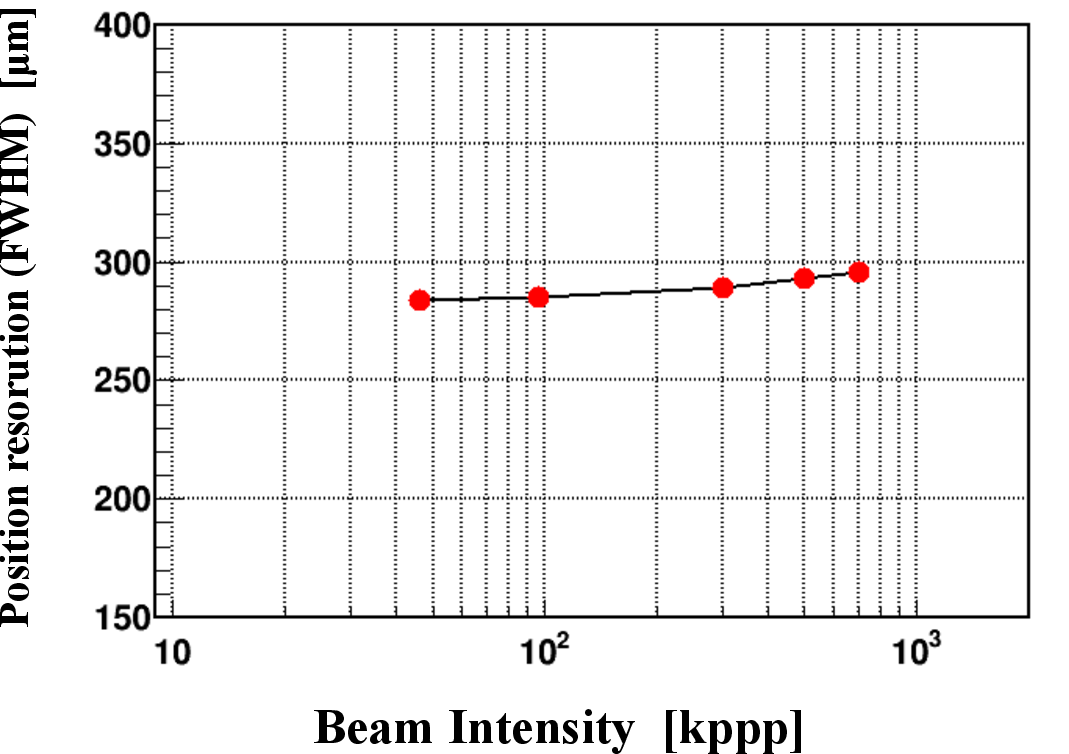}
\caption{\label{fig.posiHII}Beam intensity dependence of the position resolution. The gas pressure was 5~Torr, and the applied bias 525~V.}
\end{minipage}
\hspace{0.02\columnwidth}
\begin{minipage}{0.48\hsize}
\centering
\includegraphics[height=5cm,clip]{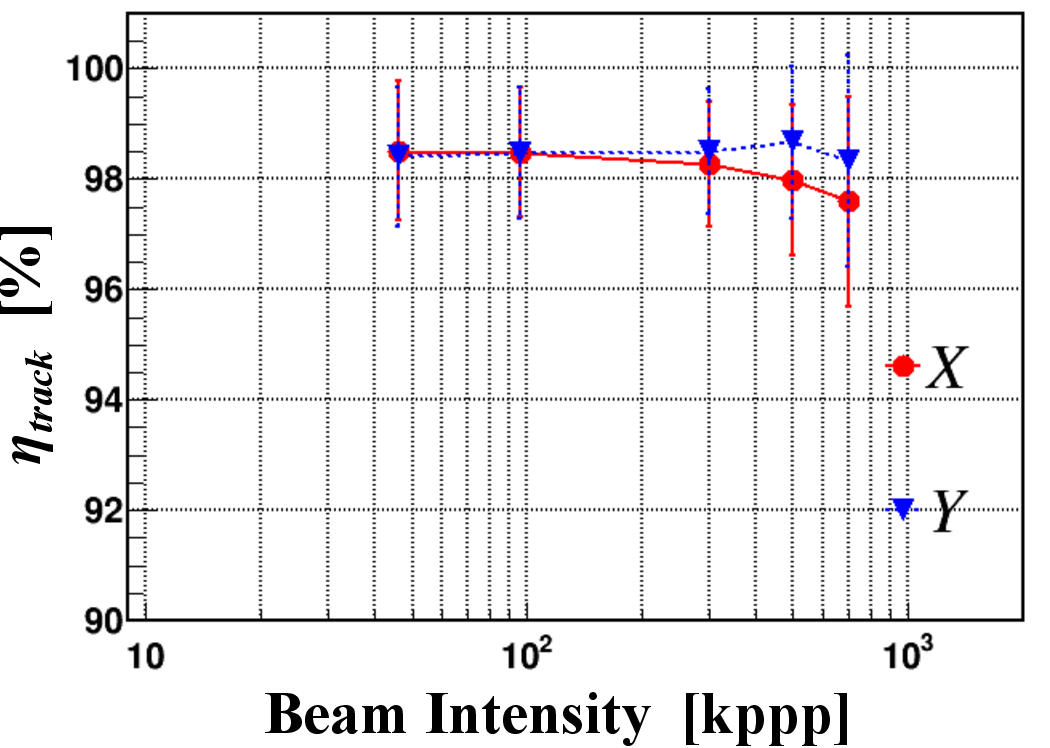}
\caption{\label{fig.trackeffint}Tracking efficiency as a function of beam intensity. The gas pressure was 5~Torr, and the applied bias 525~V.}
\end{minipage}
\end{figure}

The position resolution of SR-PPAC as a function of applied anode bias is shown in Fig.~\ref{fig.posiHI}. The bias dependence of the position resolution is similar to that of the detection efficiency of $Q_{1}$. The resolution is almost constant at the bias region where $Q_{1}$ efficiency is greater than 99\%. The best position resolution was 239~{\textmu}m (FWHM) for 10~Torr, and 285~{\textmu}m (FWHM) for 5~Torr. Figure~\ref{fig.posiHII} shows the beam intensity dependence of the position resolution. The position resolution for 700~kppp achieved 295~{\textmu}m (FWHM).

\subsection{Tracking efficiency}

The tracking efficiency was defined as follows:

\begin{equation}
\eta_{\mathrm{\mathrm{track}}} = \frac{N_{3\sigma} \otimes N_{\mathrm{mwdc}}}{N_{\mathrm{mwdc}}}
\end{equation}

\noindent where $N_{3\sigma}$ and $N_{\mathrm{mwdc}}$ represent effective number of events within 3$\sigma$ for position resolution of SRPPAC ($\sigma$ = standard deviation) and number of events in two LP-MWDCs, respectively.

Fig.~\ref{fig.trackeffint} shows the beam intensity dependence of the tracking efficiency. For a beam with an intensity of 700~kppp, tracking efficiency was 97\% for a single cathode plane.

\subsection{Position dependence of the resolution}

\begin{figure}[!h]
\centering\includegraphics[width=\hsize]{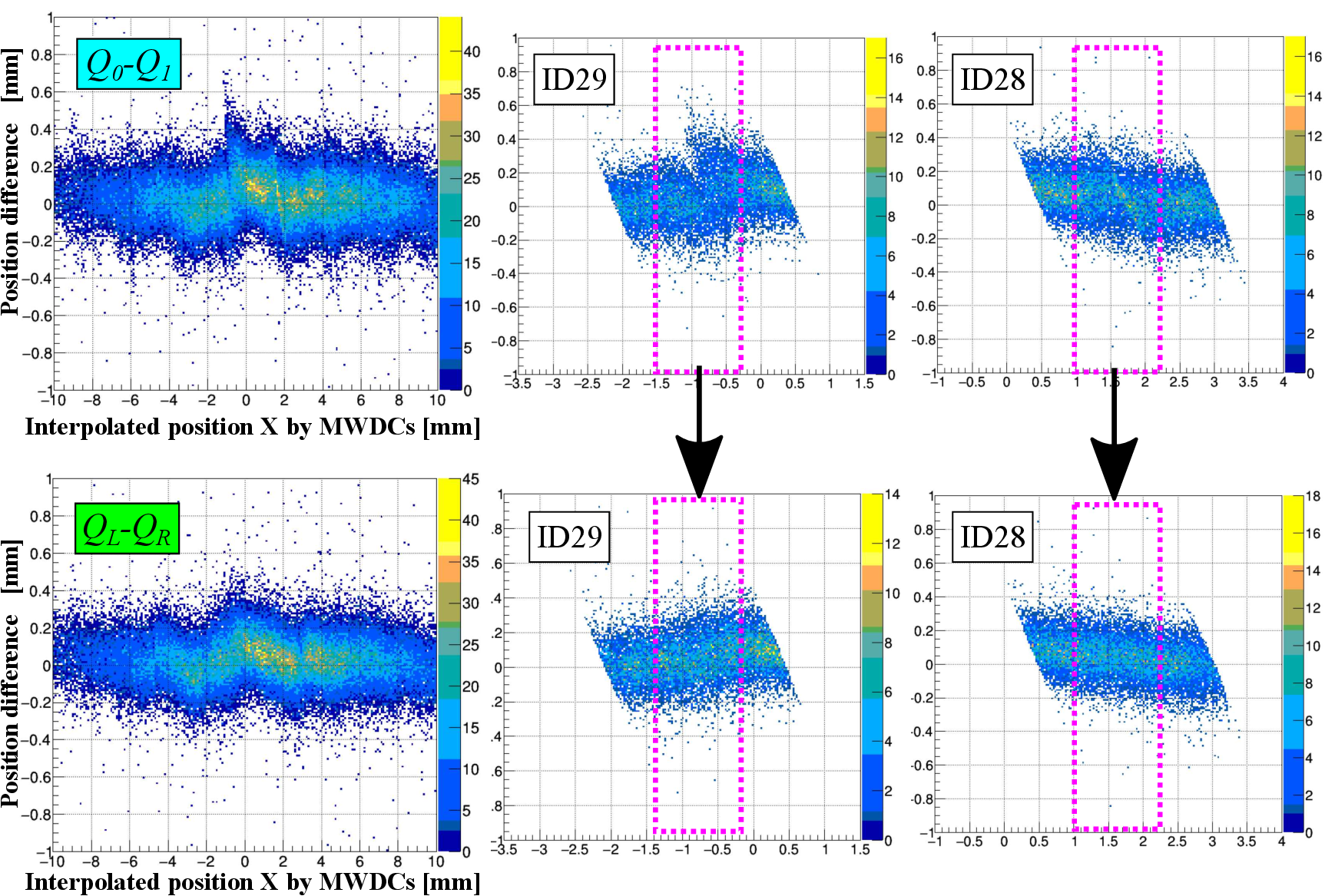}
\caption{\label{fig.posicorr}The left figures show the Position difference between SR-PPAC and Tracking of MWDCs versus tracking of MWDCs. The middle and right figures show the distribution of two strips denoted as ID28 and ID29. The top three figures are the results of using the method described in Sec.~2.  
The position resolution is poorer at the center of the strip than at the edge, and the distribution is not continuous.
The bottom three figures are the results of using the charge of the left and right side strips as described in Eq.~(8).}
\end{figure}

The average resolution was mentioned in the previous section, however the position resolution was not uniform across the entire strip width. In Fig.~\ref{fig.posicorr}, the horizontal axis represents the interpolated position by tracking using LP-MWDCs and the vertical axis represents  the position difference between SR-PPAC  and the interpolated position. The width of the distribution along the vertical axis corresponds to the position resolution. The middle and right figures are distributions of selected strips around the central axis. The top three figures are the results calculated using the method described in Sec.~2.1, which uses the difference between $Q_{0}$ and $Q_{1}$. As seen in the figure, the resolution at the center of a strip is poorer than the edge of a strip, and the distribution is not continuous. When a beam bombards close to the center of a strip, $Q_{1}$ and $Q_{2}$ are almost equal. Therefore, it is difficult to determine the true strip of $Q_{1}$ owing to the fluctuations. 

This problem can be improved by selecting proper combination of strips to take the charge difference. The charge deposited in the strip to the left (right) of the central strip shall be denoted as $Q_{L}$ ($Q_{R}$).
The charge difference between $Q_{0}$ and $Q_{L}$ ($Q_{R}$) is useful.
When a hit position goes from the right to left of a strip, the amount of charge of the strip on the left side is small and the resolution gets worse in the case that the position calclated by using $Q_{0} - Q_{L}$.
Using the difference between the charge of the central strip and the right side strip ($Q_{0} - Q_{R}$), the position dependence of the resolution is reverse.
Therefore, taking the average of position determined  using the charge of the left strip and the right strip, weighted by charge, the position dependence of the resolution will be improved. 

\begin{equation}
x = \frac{(Q_{0}-Q_{R})}{(Q_{0}-Q_{L})+(Q_{0}-Q_{R})} \times x_{L} + \frac{(Q_{0}-Q_{L})}{(Q_{0}-Q_{L})+(Q_{0}-Q_{R})} \times x_{R}
\end{equation}

\noindent where $Q_{L}$ and $Q_{R}$ represent the charge of a strip on the left and right side of strip with ID$_{0}$. $x_{L}$ and $x_{R}$ represent position calculated using $Q_{0} - Q_{L}$ and $Q_{0} - Q_{R}$, respectively. The lookup tables for calculating $x_{L}$ and $x_{R}$  were created separately for each.
The three figures at the bottom of Fig.~\ref{fig.posicorr} show the distributions calculated by this method. The position dependence is improved and overall resolution improved by approximately 10~{\textmu}m ($\sigma$). 
The futher correction to improve the position dependence of the accuracy will be required when the precise position determination is required by the specific physics case. That may  be solved by the charge calibration of ToT, which is the future study.

\subsection{Strip size dependence}

We also tested the prototype SR-PPAC with strip width of  5.14~mm to investigate the strip size dependence.
It is expected that the position resolution depends on the strip width. In this study, a PCB board to combine charge from two strips was made.

\begin{figure}[tbh]
\centering\includegraphics[scale=0.5]{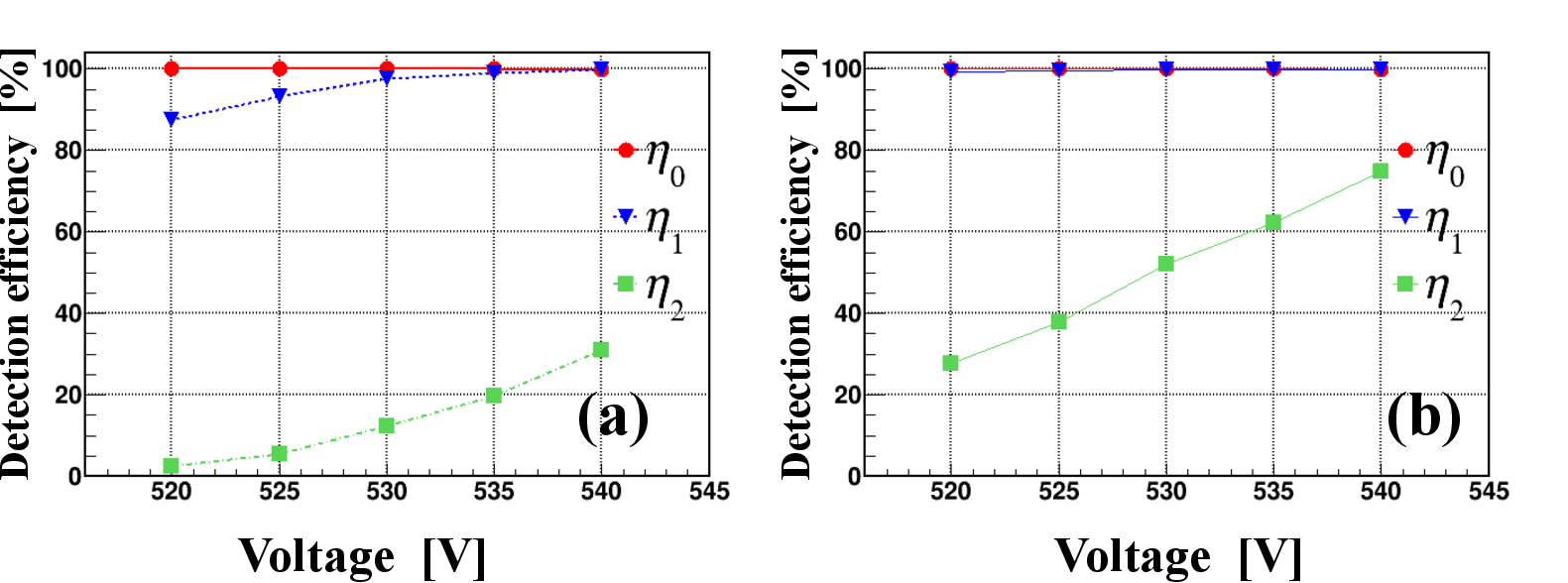}
\caption{\label{fig.effdouble}Anode bias dependence of detection efficiency of SR-PPAC of strip size of  5.14 mm operated under 5 Torr gas pressure. (a) and (b) represents X and Y respectively. }
\end{figure}

Figure~\ref{fig.effdouble} shows the detection efficiency of the strip size of  5.14~mm. The amount of charge collected from a strip depends on the strip size and the size of charge distribution on the cathode. $Q_{1}$ efficiency is enough but $Q_{2}$ efficiency is not sufficient because the amount of $Q_{2}$ is dramatically decreased in the case of 5.14~mm.

\begin{figure}[tbh]
\centering\includegraphics[scale=0.42]{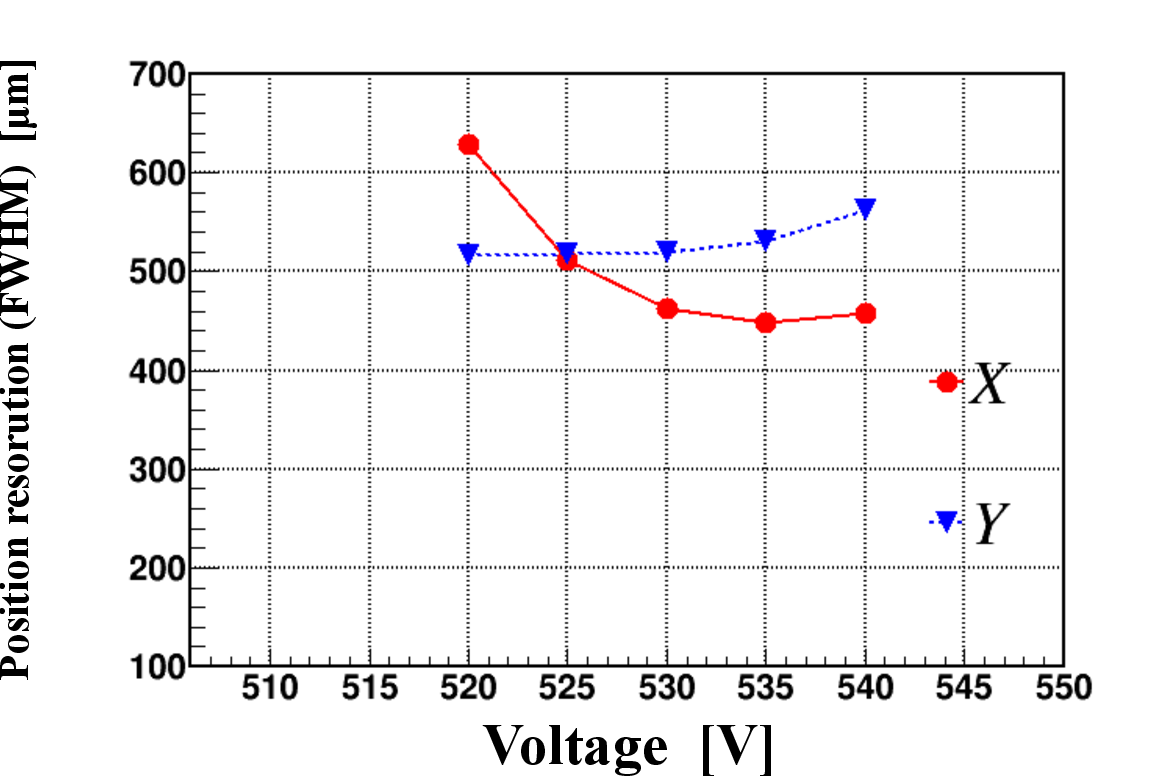}
\caption{\label{fig.posidouble}Position resolution of 5.14 mm strip width as a function of applied bias. }
\end{figure}

Figure~\ref{fig.posidouble} shows the position resolution of the strip size of 5.14~mm as a function of applied bias. The best resolution achieved was 457~{\textmu}m (FWHM) which is significantly worse than that with 2.57~mm strips. 

Considering Eq. (1), the strip size explicitly propagates the position resolution thereby the smaller strip size has the better position resolution. However, the smaller strip size has the smaller induced charge on the strip and the signal-to-noise ratio gets poorer. To optimize the strip size for the strip-readout method, a simulation study will be desirable.

\section{Test experiment for Standard SR-PPAC at RIBF}

\begin{figure}[t]
\centering\includegraphics[scale=0.5]{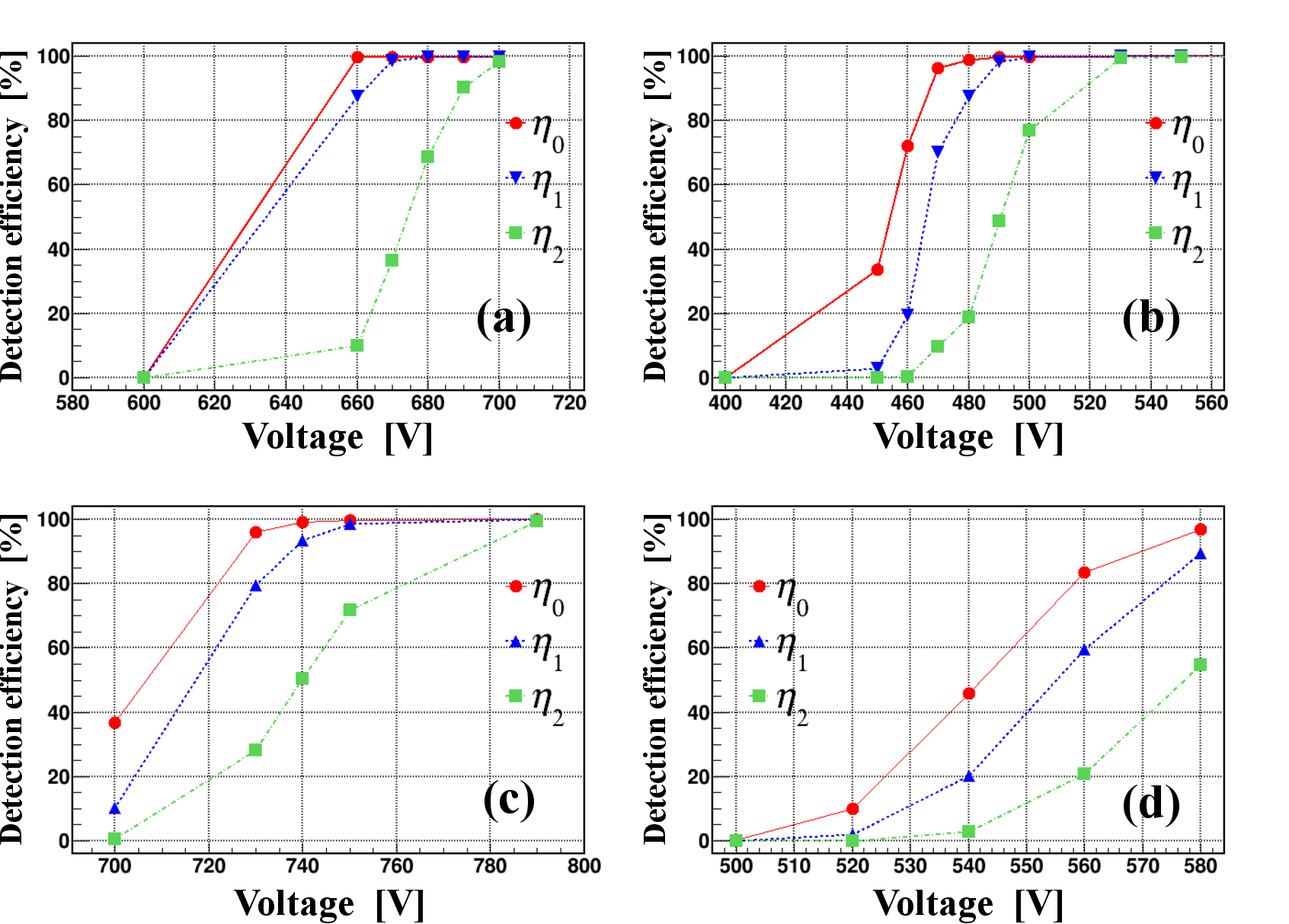}
\caption{\label{fig.effRI}Anode bias dependence of detection efficiency of  SR-PPAC downstream for a beam of 300~MeV/$u$ $^{132}$Sn, operated under gas pressure 10~Torr (a) and 4~Torr (b), and for 300~MeV/u $^{48}$Ca 10~Torr (c) and 4~Torr (d), respectively.}
\end{figure}

In this section, the performance of a standard SR-PPAC tested with continuous beams provided at the RI beam factory in RIKEN, japan was described. We implemented a standard SR-PPAC to BigRIPS separator in RIBF as a beamline detector for a physics experiment. Two SR-PPACs were installed at focal plane F5, which is a momentum-dispersive focal plane (the layout of the beamline is shown in Fig.~8 of Ref.~\cite{bigrips}). The beams were $^{132}$Sn and $^{48}$Ca with energy of approximately 300 MeV/u. Counter gas was $i-$C$_{4}$H$_{10}$ and the pressure values to be tested were 10 and 4~Torr.
For the evaluation of the detection efficiency, we used the number of events in CVD diamond detectors~\cite{dia} placed at F3 and F7 focal planes. The detection efficiency of SR-PPAC was calculated as following equations:

\begin{equation}
\eta_{i} = \frac{N_{i} \otimes N_{\mathrm{dia3}}\otimes N_{\mathrm{dia7}}}{N_{\mathrm{dia3}}\otimes N_{\mathrm{dia7}}}  \hspace{20pt} (i = 0, 1, 2)  
\end{equation}

\noindent where $N$ mean number of events detected in SR-PPAC, and $N_{\mathrm{dia3}}\otimes N_{\mathrm{dia7}}$ means the number of events in both diamond detectors at F3 and F7, respectively. The index $i$ correnponds to ID$_{0}$, ID$_{1},$ and ID$_{2}$. 
Figure~\ref{fig.effRI} shows the detection efficiency of X-plane as a function of an applied anode bias for $^{132}$Sn and $^{48}$Ca. For $^{132}$Sn, SR-PPAC can achieve more than 99\% efficiency in both 10~Torr and 4~Torr. For $^{48}$Ca the pressure of 4~Torr was not enough for desired electron multiplication. 
Applying 600~V at pressure 4~Torr, the SR-PPAC discharged. Therefore we operated SR-PPACs in 10~Torr for the $^{48}$Ca beam.

\begin{figure}[!h]
\centering\includegraphics[scale=1.25,clip]{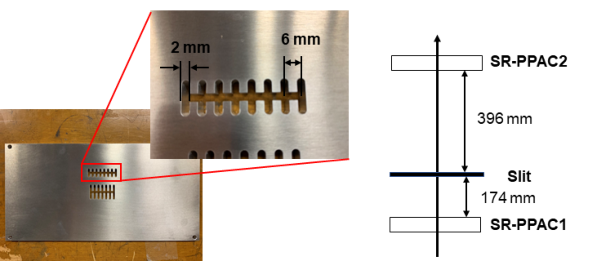}
\caption{\label{fig.fish}(Left) Photograph of fishbone slit for position deduction. (Right) illustration of the experimental setup in RIKEN for standard SR-PPAC.}
\end{figure}

\begin{figure}[!t]
\centering\includegraphics[scale=0.7]{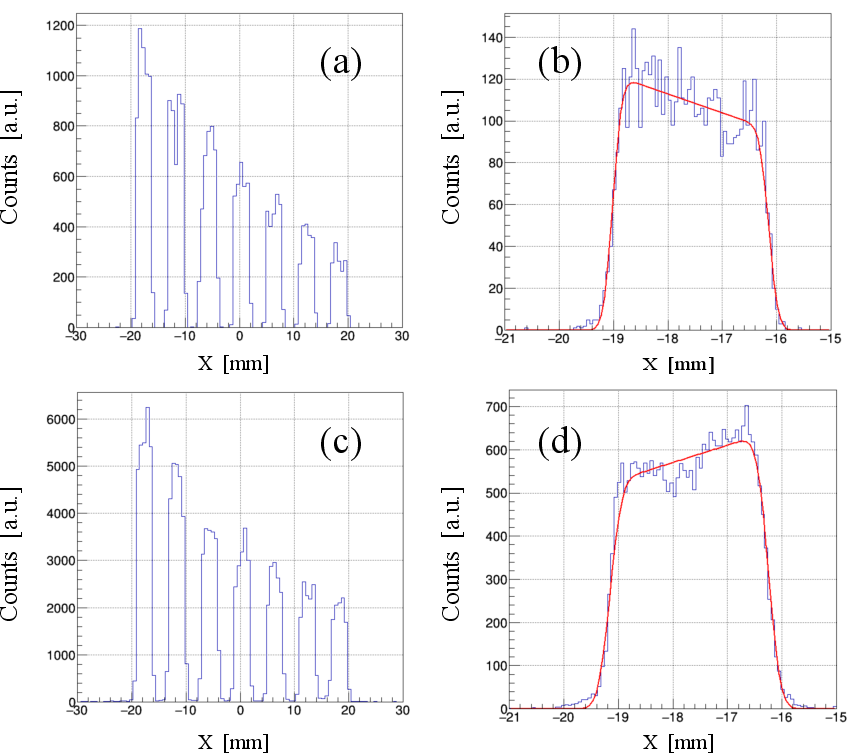}
\caption{\label{fig.posiRI}Image of fish-bone slits created by the trajectory of two SR-PPACs, for $^{132}$Sn (a) and $^{48}$Ca (c). The results of fitting of the leftmost slit for $^{132}$Sn (b) and $^{48}$Ca (d).}
\end{figure}

To evaluate the position resolution, we used a fishbone slied made of a stainless steel plate slit shown in Fig.~\ref{fig.fish} since the reference detectors cannot be installed due to the limitation by the physics experiment. The width and pitch of the slit were 2~mm and 6~mm, respectively.
An image of the slit using the beam positions detected by the two SR-PPACs was reconstructed. The beam position at the interpolated position of the slit as shown in Fig.~\ref{fig.posiRI}(a) and (c).
The sharpness of the slit image is evaluated by the fitting the error function, i.e.,  the convolution of the rectangular distribution and the gauss distribution. The position resolutions of two SR-PPACs are assumed to be same.
 
Figure~\ref{fig.posiRI}(b) and (d) show the fitting result for $^{132}$Sn and $^{48}$Ca beam distributions, respectively. The most left slit in each image has been used since it has the largest statistics. The operation bias was 515~V in 4~Torr for $^{132}$Sn, and 740~V in 10~Torr for $^{48}$Ca, which were required by the physics measurement. 
The evaluated spatial resolution was 318~{\textmu}m (FWHM) for $^{132}$Sn and 400~{\textmu}m (FWHM) for $^{48}$Ca. Considering the detection efficiency shown in Fig.~\ref{fig.effRI}, a better resolution is expected by applying 760~V for $^{48}$Ca beam.

The time resolution of  the SR-PPAC was also evaluated. PPAC can determine the timing of beams unlike MWDC because the drift time between the cathode and anode is assumed to be common for all the particles and the resulting detection time has only a common offset. The rise time of the electron signal is about 10~ns and we expected a good time resolution for SR-PPAC owing to its fast response. Since the distances between SR-PPAC and each diamond detector placed at F3 and F7 focal planes are the same, the time resolution of SR-PPAC was calculated using the difference of timing in SR-PPAC and the averaged timing of two diamond detectors as the following equation:

\begin{equation}
\sigma_{\mathrm{t\_srppac}} = \sqrt{ \sigma_{\mathrm{t\_res}}^{2} - \sigma_{\mathrm{t\_dia}}^{2} }
\end{equation}

\noindent where $\sigma_{\mathrm{t\_res}}$, $\sigma_{\mathrm{t\_srppac}}$, and $\sigma_{\mathrm{t\_dia}}$ represent the dispersion of the distribution of the difference of timings in SR-PPAC and diamond detector, the time resolution of the diamond detector, and the time resolution of SR-PPAC, respectively. We evaluated each electrode and the timing of SR-PPAC is determined as the timing of ID$_{0}$ for the cathode.
The diamond detector consists one pad on one side and four strip on the other side. In this measurment, the time resolution of each diamond detector was deduced by the difference between the timing collected from the four corners of the pad. The typical resolution of the diamond detector was 15~ps ($\sigma$).

\begin{figure}[!t]
\centering\includegraphics[scale=0.5]{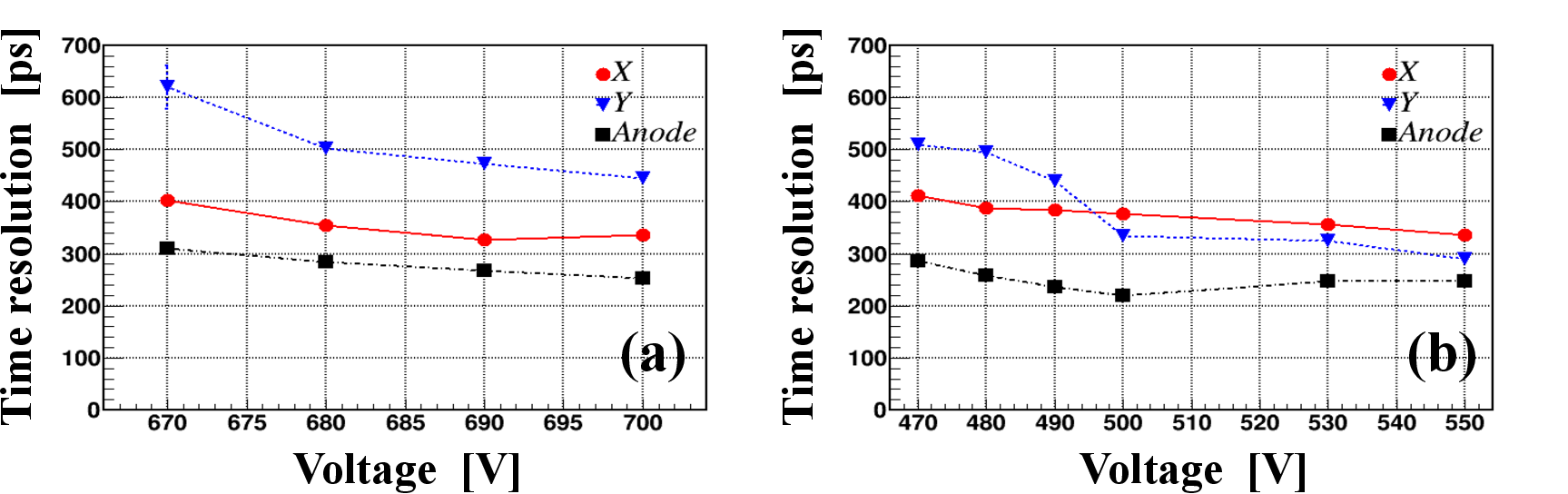}
\caption{\label{fig.timereso}Anode bias dependence of time resolution, operated under gas pressure 10~Torr (left) and 4~Torr (right) for the beam of $^{132}$Sn. }
\end{figure}

The time resolutions as a function of applied bias are shown in Fig.~\ref{fig.timereso}. When the bias is larger, the rise time of the analog signal gets sharp giving rise to the better time resolution. There is a difference in the resolution of each electrode. The time resolution of the anode plane, which has the largest deposited charge, was 260~ps ($\sigma$). 

\section{Summary and prespectives}

We have developed a strip-readout PPAC (SR-PPAC) with a fast response and a good position resolution for heavy ions. 
The SR-PPAC realizes the high detection efficiency with direct readout of charge from each strip of cathode and the charge information is acquired by using the Time-over-Threshold method. The detection efficiency was $\sim$ 99\% for 115~MeV/u $^{132}$Xe with the intensity of 700~kppp, 300~MeV/u $^{132}$Sn with 770 kHz and 300~MeV/u $^{48}$Ca with 770~kHz. 
SR-PPACs worked stably and kept high detection efficiency even for high-intensity beams with about 700~kHz over $\sim$ 120~hours. 

By using a lookup table of the charge difference between two strips to the beam position, position resolution was considerably smaller  than the strip size. The position resolution was 239~{\textmu}m (FWHM) for $^{132}$Xe with intensity of 45 kppp, and 289~{\textmu}m (FWHM) with intensity of 
700~kppp. The results indicate that the position resolution of a SR-PPAC is comparable to a LP-MWDC for heavy-ion beams.

The dependence of strip size on the position resolution will be studied using a simulation, especially for smaller strip size.
The time resolution of SR-PPAC was 260~ps ($\sigma$). Factors contributing to in the time resolution will also be evaluated by the simulation, important in accessing the timing applications of these detectors

\section*{Acknowledgment}

The work was led as part of a research project with Heavy Ions at NIRS-HIMAC. A part of the experiments was performed at RI Beam Factory operated by RIKEN Nishina Center and CNS, the University of Tokyo.
The authors acknowledge the accelerator staff at HIMAC and RIKEN for providing excellent quality beams and their stable operation of the accelerators. This work was supported by JSPS KAKENHI Grant Number JP16H06003, RIKEN Junior Research Associate Program, and Quantum Science and Technology Fellowship Program.



%

\end{document}